# Optical detection of charge defects near a graphene transistor using the Stark shift of fluorescent molecules

Carlotta Ciancico[1], Iacopo Torre[1], Bernat Terrés[1], Alvaro Moreno[1], Robert Smit[2], Kenji Watanabe[3], Takashi Taniguchi[4], Michel Orrit[2], Frank Koppens[1], Antoine Reserbat-Plantey[1,5]

[1]ICFO - Institut de Ciencies Fotoniques, The Barcelona Institute of Science and Technology, Castelldefels, Barcelona, 08860, Spain
[2]Huygens-Kamerlingh Onnes Laboratory, Leiden University, Postbus 9504, 2300 RA Leiden, The Netherlands3
[6]Research Center for Functional Materials, National Institute for Materials Science, Tsukuba, Japan4
[7]International Center for Materials Nanoarchitectonics, National Institute for Materials Science, Tsukuba, Japan
[5]Université Côte d'Azur, CNRS, CRHEA, Rue Bernard Grégory, 06905 Sophia-Antipolis, France
Corresponding authors: frank.koppens@icfo.eu ; antoine.reserbat-plantey@cnrs.fr

**Two-dimensional crystals and their heterostructures unlock access to a class of photonic devices, bringing nanophotonics from the nanometer scale down to the atomic level where quantum effects are relevant. Single-photon emitters (SPEs) are central in quantum photonics as quantum markers linked to their electrostatic, thermal, magnetic, or dielectric environment. This aspect is exciting in two-dimensional (2D) crystals and their heterostructures, where the environment can be abruptly modified through vertical stacking or lateral structuring, such as moiré or nano-patterned gates. To further develop 2D-based quantum photonic devices, there is a need for quantum markers that are capable of integration into various device geometries, and that can be read out individually, non-destructively, and without additional electrodes. Here, we show how to optically detect charge carrier accumulation using sub-GHz linewidth single-photon emitters coupled to a graphene device. We employ the single molecule Stark effect, sensitive to the electric fields generated by charge puddles, such as those at the graphene edge. The same approach enables dynamic sensing of electronic noise, and we demonstrate the optical read-out of low-frequency white noise in a biased graphene device. The approach described here can be further exploited to explore charge dynamics in 2D heterostructures using quantum emitter markers.**

The integration of a quantum sensor[1] presents a challenging dichotomy: on the one hand, the quantum probe must be isolated from its environment as much as possible to maintain its purity and coherence, while on the other hand, it must be capable of interacting with the same environment to function as a sensor. To address this challenge, solid-state quantum emitters[1] have been incorporated into various applications such as nanoscale thermometry[2], charge sensing[3], and magnetometry[4]. One limitation of quantum sensing that utilizes dispersive schemes based on measuring emission line shifts is the presence of inhomogeneous broadening. This phenomenon increases the emission linewidth, compromising the emitter's coherence and ultimately limiting the accuracy of the spectroscopic detection of the emitter. In quantum sensing, another persistent challenge is to position the quantum marker as close as possible to the target sample. This proximity is essential for reducing the probed volume to the nanoscale and, ultimately, to the level of a single particle. This proximity is essential for reaching quantum sensing applications' high sensitivity and resolution. To address these limitations, there is a need for narrow linewidth quantum emitters that are both bright and stable and can be positioned at the nanoscale.

In the field of quantum nanophotonics, nanoscale quantum markers are of great significance, mainly when used in devices based on two-dimensional materials (2DM)[3]. These materials offer atomic-level control of light-matter interactions at the nanoscale[3,5], which is promising for tailoring quantum light-matter interactions. In these 2DM devices, the abrupt variation of material properties in van der Waals stacks – where metal, insulator, ferromagnets, and semiconductors can be stacked on top of each other, one atomic layer after another – requires highly sensitive and selective probes. Various efforts have been made to embed such emitters to meet the demands of individual control and integration of single-photon emitters within 2D materials directly. These efforts include using local strain fields[3,6–8], defects implantation[9,10], nano-indentation[3], and even moiré-confined excitons in twisted heterostructures[3]. However, such SPES generally perform worse than other solid-state quantum emitters[1] in terms of linewidth[3] ($\Gamma/2\pi \sim 40$ GHz), emission intensity[8] ($I_{sat} \sim 10^4$ counts/s), and sequential single-photon emission[11] ($g^{(2)}(0) \sim 0.2$). In this work, we use a second approach of building a hybrid device made of nanoscopic sub-GHz linewidth solid-state SPES integrated with a 2D material[3,12]. Among the available solid-state SPES, aromatic molecules[3] such as dibenzoterrylene (DBT) are suited for integration at the nanoscale as the synthesis of DBT-doped anthracene nanocrystals (NCs) is well controlled[13]. This unlocks the design of hybrid systems combining atomically thin electronic interfaces with SPES with narrow linewidth ($\Gamma/2\pi \sim 50$ MHz) at low temperatures, high photon detection rates ($I_{sat} \sim 10^6$ counts/s) and almost no multi-photon events ($g^{(2)}(0) < 0.03$).

Previous studies have demonstrated that these aromatic molecules can be used in quantum technology applications, where properties such as emission decay rate[3] and energy levels[3,14] can be tuned. In particular, these molecules serve as markers or probes, revealing environmental modifications induced by near-field interactions in the vicinity of the 2D material[3,5,15,16]. The advantage of such a large sensitivity is that emitters can be used to probe local fluctuations of the electric field[17,18]. Local probing of graphene devices has led to observing the electron-phonon Cherenkov effect[3], mapping the current flow[3], and probing the electronic scattering at the edge[19] or local compressibility[20]. Single molecules have also been predicted as a probe of nanomechanical oscillations of carbon nanotubes[21]. The unique electrostatic environments provided by 2D materials, such as semimetallic graphene and semiconductor transition metal dichalcogenides, can be identified through the emission spectra of molecules[3]. In particular, certain 2D materials, like hBN, can stabilize molecular emissions[12]. Furthermore, in various graphene-related applications, including gas sensing, metrology, and quantum transport, electronic noise – especially at low frequencies – poses a challenge for high-precision measurements[22,23]. There is a need for sensitive, large bandwidth and integrated sensors of electronic noise to provide insights into the local origin of noise. Combining a 2D material with DBT molecules appears as an original platform, where the quantum emitters can also act as probes of the electric fields near the 2D material[3,5,15,16].

Our device combines a gate-tunable graphene structure and sub-GHz linewidth quantum emitters. A stack made of single-layer graphene on top of a hBN multilayer ($\sim 50$ nm thick) is deposited on a $SiO_2/Si^{--}$ chip and connected to source and drain gold electrodes (see Figure 1a). We iteratively spincoat a polymeric suspension doped with nanocrystals of anthracene containing DBT molecules until anchoring some at the center and near the edge of the graphene device (see Figure 1b,c). DBT molecules are solid-state quantum emitters that can be modeled as two-level systems at cryogenic temperatures. We excite resonantly the zero-phonon line transition (00ZPL) from the zero-vibrational electronic ground state to the excited state using a tunable 785 nm CW laser source. The molecule decays to vibrational ground states (phonon sideband), and we detect these red-shifted photons *via* a single-photon counting module, filtering out the elastically scattered photons. Upon scanning the excitation energy, we measure a typical forest of narrow peaks (see Figure 1d), each corresponding to an individual molecule. The emission frequencies of the peaks are spread over hundreds of GHz due to local variations of strain and charge distribution in the crystalline environment. By integrating the emission signal for each position of the laser, we obtain a high-signal-to-noise fluorescence map that reveals the position of nanocrystals (see Figure 1c). The resolution of our confocal microscope does not reveal the individual position of single molecules inside a nanocrystal. Still, it is sufficient to locate the nanocrystals at the center or the edge of the graphene device with a spatial resolution of about 700 nm. For each nanocrystal, we observe around a hundred lines corresponding to individual molecules (see Figure 1d). At the center of the device, the linewidth distribution is broad, peaked at $\Gamma/2\pi \sim 200$ MHz (see inset of Figure 1d), often attributed to non-resonant energy transfer to graphene[3]. It is possible to reduce this effect and maintain Fourier-limited linewidths by using hBN as a substrate for the DBT:Ac crystal[12]. Our geometry would require a top hBN encapsulation of the graphene device, thus adding an extra distance from the graphene to the molecule. The studied device shows two clusters of nano-particles containing DBT molecules, one at the center and one at the edge (Figure 1b,c).

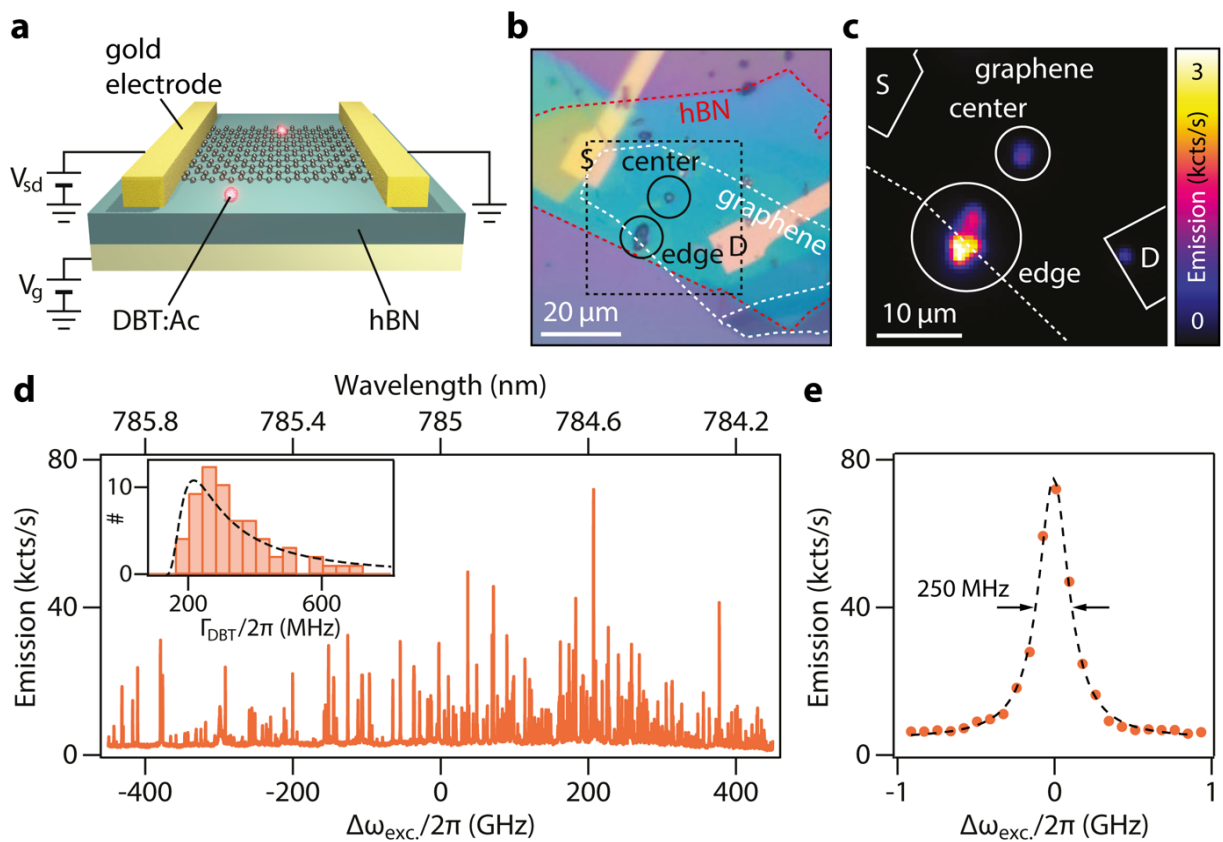

**Figure 1 – Hybrid photonic system integrating single organic molecules into a graphene device. (a)** sketch of the graphene device and electronics schematics. Nanocrystals doped with single molecules (DBT:Ac) are deposited over the sample. In the studied device, one lies at the center and another at the edge of the graphene device. **(b)** Optical microscope image of the graphene device with source

(S) and drain (D) electrodes. Doped nanocrystals deposited respectively at the center and at the edge of the graphene flake are highlighted with a black circle. **(c)** Fluorescence map of the dashed area in **(b)** obtained by resonant confocal excitation. **(d)** Typical emission spectrum of one doped nanocrystal. The inset shows a histogram of linewidths for more than 60 molecules. **(e)** Zoom on one molecule's emission with linewidth $\Gamma_{DBT}/2\pi \sim 250$ MHz. Measurements shown in c,d and e are taken at $T = 2.8$ K, using a confocal spot of 700 nm and with all source, drain and backgate electrodes grounded.

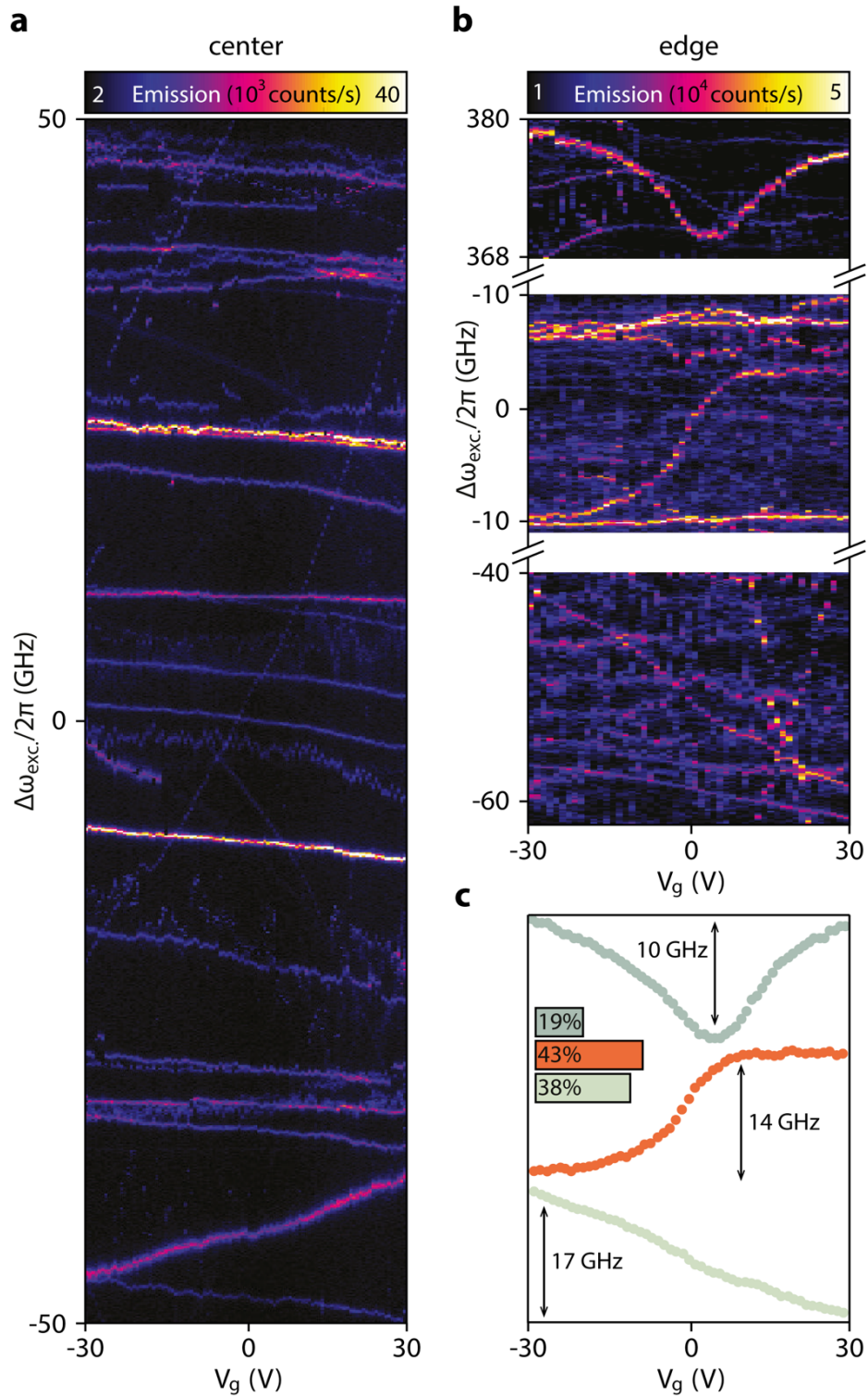

**Figure 2 – Exotic Stark shift response at the edge of the graphene device (a)** For more than 80 molecules measured in the nanocrystal at the center of graphene we observe a close-to-linear Stark shift. **(b)** Molecules in the nanocrystal positioned at the edge of graphene show a different response in the frequency shift as a function of V_g. **(c)** For more than 180 molecules detected at the edge, approximately 38% exhibit linear Stark shift, while 43% and 19% show an atypical "Z" and "V" behavior, respectively.

To show how single DBT molecules can serve as local probes of their electrostatic environment, we measure the shift in their emission frequency as a function of the applied back-gate potential $V_g$ at these two remarkable positions – center and edge – of the graphene device. In this experiment, we consider that the exchange of electrons between the molecule and the graphene is negligible; hence, the molecules are not measuring the electrochemical potential but strictly the electric field. When positioning the excitation laser at the position of the nanocrystal located at the center, spectra of molecules are recorded for gate-voltage values of $V_g \in [-30; 30\text{V}]$ (Figure 2a). Each line corresponds to one molecule within the nanocrystal. For each molecule measured at the center position, we generally observe a monotonic variation of the frequency shift $\delta\omega(V_g)/(2\pi)$ with backgate voltage $V_g$, which is roughly linear over the whole span of $V_g$ and ranging from 10 MHz/V for the almost insensitive to 1 GHz/V for the most sensitive case. Most molecules show stable emission in time ($\tau_{meas} \sim 6$ h). The monotonic Stark behavior measured on DBT molecules located at the center of the graphene device is similar to what has been previously reported under an in-plane electric field[24].

In contrast, when illuminating the nanocrystal at the edge of graphene, we sometimes observe a saturation of $\delta\omega(V_g)$ upon increasing $V_g$ (Figure 2b). In most cases, we observe that $\delta\omega(V_g)$ varies either linearly or quadratically ("Z" and "V" behaviors in Figure 2b) in the range $V_g \in [-15; 15\text{V}]$. For "Z" behaviors, it reaches a clear saturation plateau for $|V_g| > 15$ V, while less pronounced for "V" cases. Such saturating behavior is observed for approximately 2/3 of the detected molecules near the graphene's edge ($\sim 1/3$ shows a linear dependence, see Figure 2b bottom). Due to long acquisition time ($\delta t \sim 600$ s) per spectra and small voltage step ($\delta V_g \sim 1$ V), the resulting small voltage sweep rate ($\delta V_g/\delta t \sim 1.6$ mV/s) prevents any hysteretic behavior. We have performed more than ten sweeps in both directions and systematically observed similar behavior.

One of the main mechanisms responsible for the frequency shift of the molecules $\hbar\delta\omega$ is the interaction of the molecule with the external electric field, known as the Stark effect[3,24,25]. Other mechanisms, such as strain or thermal variations, can be reasonably ruled out as the device is not piezoelectric, the laser power is kept constant[26] and no Joule dissipation is expected when varying $V_g$. Molecular Stark shift can be explained with a contribution coming from the interaction of the dipole change between ground and excited state ($\delta\boldsymbol{\mu}$) and the corresponding polarizability change ($\delta\bar{\bar{\alpha}}$) with the external electric field $\boldsymbol{E}$, following the formula:

$$\hbar\delta\omega = -\delta\boldsymbol{\mu} \cdot \boldsymbol{E} - \frac{1}{2}\boldsymbol{E} \cdot \delta\bar{\bar{\alpha}} \cdot \boldsymbol{E}. \quad (1)$$

At the center of the device, the electric field scales linearly with the voltage $V_g$ applied between the two electrodes. This leads to a parabolic dependence of the energy shift with coefficients that depend on the position $\boldsymbol{r}$ and orientation of the molecule[3]. This is sufficient to describe qualitatively the behaviors we observe at the center position (see Figure 2a), where only small fields generated by defects, charge inhomogeneities[27] or stray fields are present.

At the edge of the graphene device (see Figure 2b), the electric field is stronger and has an in-plane component, as shown in the finite-element simulation of Figure 3c. To explain the saturation behaviors observed at the edge, the previous model needs to be refined. The energy saturation suggests that certain charges can get trapped and become insensitive to the backgate voltage, thus leading to a saturating contribution in the total electric field $\boldsymbol{E}$. To account for this behavior, we develop a minimal model of a defect based on the following five assumptions i) the defect is treated as a small piece of conductor of size $R$ (see Figure 3c); ii) the defect is in thermodynamical equilibrium with the graphene, iii) the electrons trapped inside the defect can only occupy one energy level with energy $\epsilon_D$ and degeneracy $g$, and iv) the Fermi-Dirac distribution at the same chemical potential and temperature of graphene controls the occupation of the defect. Finally, v) the interchange of charge between the defect and graphene happens on timescales faster than the time it takes for a molecule to emit a photon $\Gamma_{\text{DBT}}^{-1} \sim 0.64$ ns.

As a result, the number of trapped electrons in such a defect can vary only from 0 to $g$, thus providing a saturation mechanism. The Stark shift of the molecule emission is determined by the number of charges present on the defect and should in principle, be quantized. However, if the charge fluctuation dynamics is faster than the molecule's lifetime this quantization effect is averaged out and the Stark shift is sensitive only to the average occupation of the defect.

Thanks to the linearity of electrostatics, the electric field at the position of a molecule depends linearly on both $V_g$ and the electrostatic potential drop between the gate and the defect – $\phi_{gD}$ – via geometric coefficients (See details in Supplementary Information Sect. 1). However, this potential $\phi_{gD}[V_g]$ is a non-linear function of $V_g$ thanks to the finite electronic occupation of the defect. According to our toy model (See full derivation in Supplementary Information Sect. 2), we find that within the experimental conditions ($T = 2.8$ K), $\phi_{gD}[V_g]$ is well approximated by a piecewise linear function of $V_g$. It is equal to $V_g$ around zero: $\phi_{gD} = V_g$ for $V_g \in [V_{g,1}; V_{g,2}]$), and becomes practically constant outside this range, as illustrated in Figure 3b. Effectively, the non-linear relation between $\phi_{gD}$ and $V_g$ can be represented by an equivalent circuit in the form of a voltage clipper – more rigorously, a Zener diode clipper – with thresholds $V_{g,1}$ and $V_{g,2}$ as shown in Figure 3a.

The Stark shift for a molecule located at the edge, taking into account the non-linear response of the defect, can be expressed as

$$\hbar\delta\omega = a\,V_g + bV_g^2 + a'\phi_{gD}[V_g] + b'\,V_g\phi_{gD}[V_g] + b''\bigl(\phi_{gD}[V_g]\bigr)^2, \qquad (2)$$

where $a, b, a', b', b''$ are coefficients depending on the molecule's position and orientation (see derivation in Supplementary Information Sect. 1). The coefficients $a, a'$ are proportional to $\delta\boldsymbol{\mu}$, and they stand for the linear contribution to the Stark shift. On the other hand, the coefficients $b, b', b''$ are proportional to $\delta\bar{\bar{\alpha}}$ and are responsible for the quadratic dependence with $V_g$, for small values of $|V_g|$ (before the saturation occurs). The last three terms of equation (2) describe the saturating behavior and are graphically depicted in Figure 3b. The molecules showing a linear shift before saturation can be mainly described with the term $a'\phi_{gD}[V_g]$, while for the quadratic response, a combination of the three terms is required.

We extracted the following quantities from the experimental results: i) the span in backgate voltage $\Delta V_g$ before saturation occurs, and ii) the difference in the molecule's frequency from one saturation region to the other $\hbar\Delta\omega_0 = \hbar\delta\omega(V_{g,2}) - \hbar\delta\omega(V_{g,1})$. The first quantity can be expressed as a function of the unit charge $e$, the state degeneracy $g$ and the defect self-capacitance $C_D$ (see Supplementary Information Sect. 2):

$$\Delta V_g = V_{g,2} - V_{g,1} = eg/C_D . \qquad (3)$$

These quantities $\Delta V_g$ and $\hbar\Delta\omega_0$ are shown in Figure 3d for 21 measured molecules located close to the graphene edge. A spread of $\Delta\omega_0/2\pi \sim 300$ MHz is observed for all the molecules, independently of the linear or quadratic response. This shift in frequency is equally distributed around positive and negative values supporting the hypothesis of random induced dipole's orientation. In contrast, $\Delta V_g$ shows a narrower distribution centered around $V_g = 15$ V. The spread is contained within 5 V, thus suggesting that all the molecules are probing the same type of defect.

To prove that our results are compatible with a defect of atomic size with degeneracy close to unity, we are going to provide an estimation of the defect size $R$ and the degeneracy $g$. From this experimental observation of $\Delta V_g$, replacing the capacitance of the defect expressed as $C_D = 4\pi\epsilon_0 R$ in Equation (3), we can estimate the ratio $R/g = e/(4\pi\epsilon_0 \Delta V_g) \sim 10^{-10}$ m. From such a simple estimation, we obtain the lower bound for the defect size. The minimum value of the degeneracy g being 1, the defect size $R$ cannot be smaller than an atom, serving as a sanity check of our estimation. Of course, this ratio cannot rule out the possibility of a more significant defect with higher degeneracy $g \gg 1$. To get more insight into the defect's size, we perform another independent estimation of $R$. Here, we consider that the full span of frequency shift $\overline{\hbar\Delta\omega} = \max(\hbar\Delta\omega) - \min(\hbar\Delta\omega)$ covers all dipole orientations over the full solid angle. This hypothesis implies that there is no preferential direction in the dipole orientation distribution and that the 21 measurements shown in Fig. 3d accurately sample this distribution.

We also consider a spherical defect for simplicity located at a distance $r$ to the molecule. The upper bound of $r$ is given by the excitation laser spot size ($\emptyset \sim 700$ nm). This estimation leads to $R \cong \overline{\hbar\Delta\omega}\, r^2/(2\,\Delta V_g\,|\delta\boldsymbol{\mu}|) < 5$ Å, assuming $\overline{\hbar\Delta\omega} = 0.3$ GHz, $r < 700$ nm, $\Delta V_g = 15$ V ; $|\delta\boldsymbol{\mu}| = 2$ D (from ref[3]). These assumptions reinforce the idea of an atomic scale defect with possible degeneracy $g$ spreading from 1 to 5. Nevertheless, our analysis is limited when it comes to identify the exact nature of the defect. Such defects can be located near the edges of graphene, supporting previous interpretations in transport measurements of graphene nanoribbons[28]. Nonetheless, contamination—potentially introduced during the stacking process—may also reside on top of the device, acting as a conductive charge defect coupled to the graphene layer. Our main conclusion is that we can probe optically atomic-size charge defects using fluorescent molecules as local markers of the electromagnetic environment.

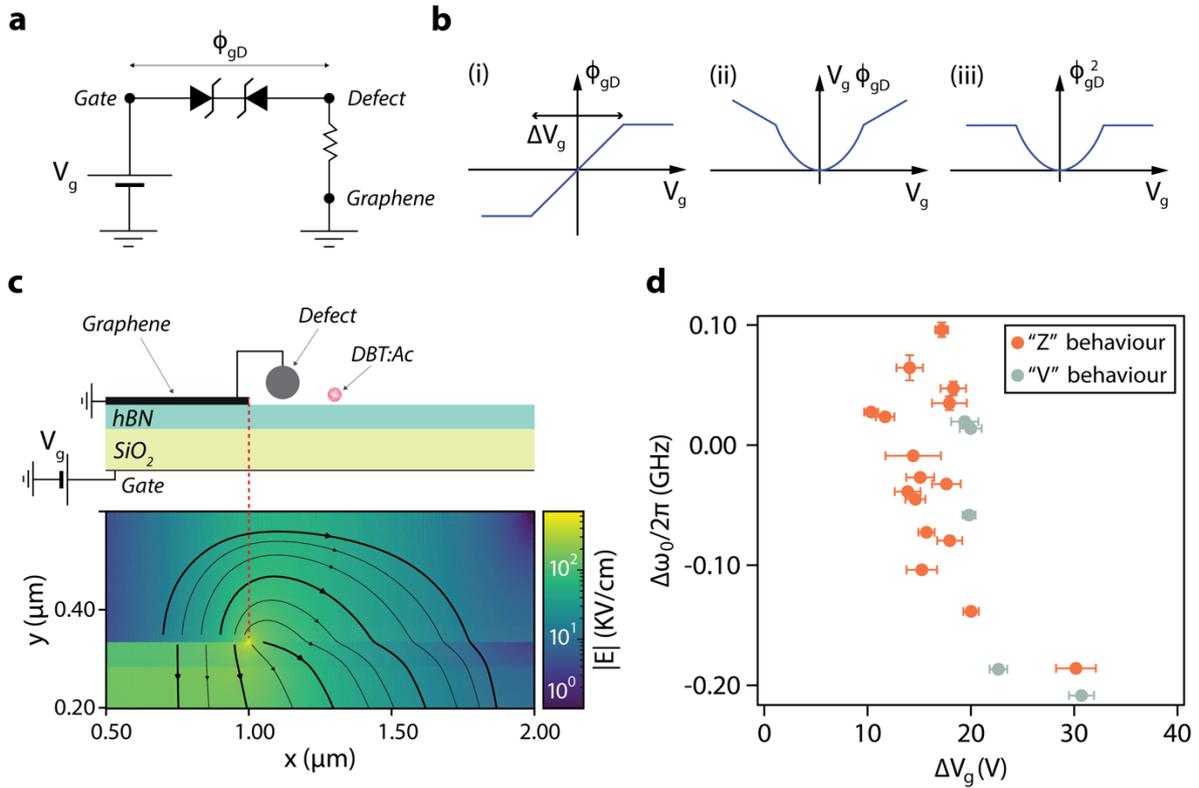

**Figure 3 – Sensing electronic defect states in graphene with sub-GHz linewidth emitters. (a)** Electrical equivalent circuit diagram of the sample. **(b)** Graphic representation of the contributions of the charge defect potential ($\phi_{gD}$) into the Stark shift equation (last three terms of Equation 2). **(c)** Side view of the graphene edge (center) and finite element electrostatic simulation of the electric field intensity (bottom). The black lines are the field lines. In our model, a defect is electrically connected to the edge of graphene and induces a modification of the electric field sensed by a doped nanocrystal nearby. The simulation does not include the perturbation due to the nano-crystal dielectric environment. **(d)** Difference in the molecule's frequency from one saturation region to the other ($\Delta\omega_0/2\pi$) with respect to the corresponding span in backgate voltage ($\Delta V_g$) extracted for 21 molecules (5 "V" and 15 "Z" behaviors) within the same nano-crystal.

In an adiabatic regime, as discussed until now, the amount of electronic charges is constant in time. We now turn to the non-adiabatic regime, which is of interest for optical sensing of electronic noise or charge fluctuation in general[18]. Considering the molecule's short lifetime of the order of nanoseconds, we assume the Stark shift to be instantaneous compared to the measurement time of the order of seconds. We collect temporal evolutions of the emission spectra for a molecule located at the center of the graphene device for different bias voltages $V_{sd}$, while keeping the backgate potential constant at $V_g = -30$ V. The situation differs from the previously discussed one; molecules at the center are not expected to be affected by the charging of a defect near the edge of the device. We observe spectral fluctuations of the emission energy in time, which are more pronounced at larger bias voltages (see Figure 4a,b).

We fit each emission line with a Lorentzian function and extract time traces for the emission frequency $\delta\omega/2\pi$ and the linewidth $\Gamma$. The main fluctuating quantity is the emission frequency $\delta\omega/2\pi$, which we characterize by computing the power spectral density ($PSD$), shown in Figure 4c. A typical $1/f$ behavior is generally present at low frequencies, while the DBT emission energy noise becomes frequency-independent (white noise) at higher frequencies (typically $> 10^{-2}$ Hz). The $1/f$ behavior here can be due to long-term laser wavelength noise caused by temperature or current fluctuations. Another contribution of such low-frequency white noise arises from current intensity fluctuations in the graphene device. Indeed, when increasing $V_{sd}$ from 0 to 20 mV, we observe an enhancement of the white-noise over two orders of magnitude (see Figure 4c). For larger $V_{sd}$, the electrical current in graphene increases, leading to enhanced current intensity fluctuations. As intensity fluctuations depend on mobility and charge fluctuations[22], we expect a similar effect as introduced above: charge fluctuations near the DBT molecule induce bias-dependent emission lineshift fluctuations, as shown in Figure 4a,b.

Current intensity fluctuations should also affect the measured linewidth, either due to a local heating effect or to inhomogeneous broadening caused by higher frequency noise. To verify this, we compute the histograms of the DBT linewidth Γ measured in time and extract the mean value $\bar{\Gamma}$ for each source-drain bias $V_{sd}$ (Figure 4d). The central values of such histograms increase from $\bar{\Gamma}(0 \text{ mV}) = 122$ MHz to values above 220 MHz at higher $V_{sd}$ values. In the presence of electric field fluctuations, we expect the Stark effect to cause inhomogeneous broadening of the measured emission line, as the measurement time (~ 3 s) is much slower than typical electronic fluctuations. Thermal contributions due to the dissipated Joule power in the graphene device could also lead to a homogeneous linewidth broadening[17,29]. In such a case, the local temperature $T_{loc}$ is directly proportional to the Joule dissipated power $P_{diss} \propto V_{sd}^2$, the linewidth broadening typically follows an Arrhenius law of the form $\bar{\Gamma} = \Gamma_0 + A \exp[-\Delta E/(k_B T + a V_{sd}^2)]$, where $\Gamma_0$ is intrinsic linewidth at zero bias, $A, a$ are constants having the unit of a linewidth and J.V$^{-2}$, respectively. $T$ is the local temperature and $\Delta E$ is a parameter with the dimension of an energy. In our experiment, the increase of $\bar{\Gamma}$ would correspond to only a few kelvins change in the local temperature[29]. Moreover, any static change in the local temperature might also lead to a strain-induced shift in the DBT emission frequency $\delta\omega/(2\pi)$ caused by the thermal expansion of the anthracene crystal[30], which we do not observe. Therefore, we argue that the variation of the mean linewidth $\bar{\Gamma}$ and the noise level for DBT emission energy observed in Figure 4c,d are most likely due to local electronic noise fluctuations when increasing the bias voltage.

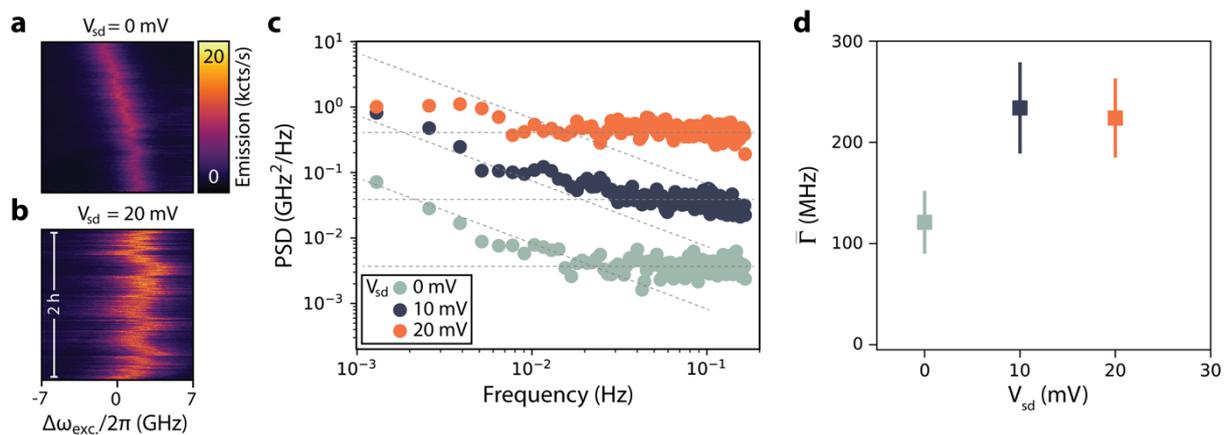

**Figure 4 – Optical readout of the electronic noise in a graphene device.** Time evolution of the same DBT molecule's spectrum at 0 **(a)** and 20 mV **(b)** bias voltage applied to the graphene device. The DBT molecule measured here is embedded in a nano-crystal located at the center of the graphene device, and its Stark shift is linear. Acquisition time for each spectrum is 3 seconds and $V_g = -30$ V. Each spectrum is fitted with a Lorentzian to extract the central emission frequency ($\delta\omega/2\pi$) for the spectral fluctuations analysis. **(c)** Power spectral density (PSD) of $\delta\omega/(2\pi)$ as a function of the frequency for the same molecule at different bias voltage $V_{sd}$. The dashed lines are guidelines for pink noise (1/f) and white noise (constant). **(d)** Average emission linewidth $\bar{\Gamma}$ as a function of the bias voltage $V_{sd}$. Error bars show the width of the $\bar{\Gamma}$. All measurements are performed at $T = 2.8$ K.

In conclusion, we have demonstrated the utilization of sub-GHz linewidth single-photon emitters (DBT molecules) as markers to probe charge defects near the edge of a graphene transistor. We have reported saturating Stark behavior when the emitters are located at the edge of graphene. Notably, the exact nature of such defects remains uncertain as we cannot distinguish a structural defect at the graphene edge from an impurity on top of graphene or a bubble at the interface acting as a reservoir of electrons. Still, our observations and model are consistent with electronic defects' charging dynamics in the unitary limit degeneracy ($g \in [1; 5]$). The optical approach developed here can be further associated with transport or scanning probe measurements to gain insights into the role of graphene edges in nanostructures[28,31]. Finally, when located at the center of the graphene device, we have reported an increase in the noise of the SPE's resonance frequency and a broadening of their linewidth upon applying a source-drain bias, which is compatible with an increase of the local electronic fluctuations sensed by the SPE. The approach described here can complement previously reported techniques to map the charge carrier density in graphene devices using NV center scanning magnetometry[32] or hBN-defect as local markers[33] directly embedded in 2D heterostructures. It could also be helpful for the quantum sensing architecture of exotic electronic phases in 2D materials (hydrodynamic regimes[34], electronic crystals[3,35]).

**ACKNOWLEDGMENTS**

This project has received funding from the EraNET Cofund Initiatives QuantERA under the European Union's Horizon 2020 research and innovation program grant agreement No. 731473 (Project acronym: ORQUID). C.C. acknowledges financial support by the ICFOstepstone funded by the Marie Skłodowska-Curie Co-funding programs (GA665884). We also acknowledge financial support from the Spanish Ministry of Economy and Competitiveness (MINE- CO), through the "Severo Ochoa" Program for Centers of Excellence in R&D (SEV-2015-0522 and SEV-2015-0496), support by Fundaciò Cellex Barcelona, Generalitat de Catalunya through the CERCA program. This work received funding from the European Union's Horizon 2020 research and innovation program Quantum Flagship (Grant No. 820378). Funded by the Agency for Management of University and Research Grants (AGAUR) 2017 SGR 1656. A. R-P thanks UCA-JEDI (ANR-15-IDEX-01), Université Côte d'Azur (CSI-2023), ANR NEAR-2D and Doeblin Federation (FR2800). R. S. and M. O. acknowledge financial support from NWO (Spinoza Prize Orrit).

**METHODS**

**Fabrication of the gate-tunable graphene device.** The hexagonal boron nitride (hBN) and graphene flakes are obtained by mechanical exfoliation of hBN and HOPG crystals, respectively. The van der Waals heterostructurs[36] made of bottom hBN and graphene is transferred onto an n-doped $Si$ wafer with a 285 nm layer of thermally grown SiO$_2$ on top. The bottom hBN has a thickness of $\sim$ 50 nm. The source and drain electrodes are designed *via* electron-beam lithography and contacted by thermal deposition of 50 nm of gold.

**Preparation of the DBT:Ac nanocrystals.** A suspension of poly(vinyl alcohol) (PVA) doped with DBT molecules hosted into anthracene nanocrystals is prepared by reprecipitation. The suspension is spin-cast onto the bottom hBN-graphene stack.

**Optical setup.** Measurements are performed at cryogenic temperatures (2.8 K), using a Montana closed-cycle cryostat. A custom-built confocal microscope is used to isolate the signal of single molecules from the background and to spatially confine the excitation and detection to a small volume. The continuous wave (CW) laser used, a tunable 785 nm diode laser (Toptica DL 100 DFB), is focused into the cryostat with a long (10 mm) working distance microscope objective (100x Mitutoyo Plan Apo NIR HR Infinity Corrected) with a numerical aperture $NA = 0.7$ and transmission approximately 60-80 %. The objective is mounted on a 3D piezo stage. A point-like detection is obtained using a single-mode fiber coupled to an avalanche photodiode (APD).

**Optical detection and electrical device actuation.** Single molecules are excited at 785 nm (5 nW/$\mu m^2$). The red-shifted single-photon emission is spectrally isolated with a long-pass filter and measured with a single-photon counting module. Fluorescence maps are collected by artificially broadening the laser excitation energy to simultaneously excite many molecules within the same excitation spot. In this case, the diode laser current is modulated at a frequency $f = 200$ Hz, which is faster than the single-point acquisition frequency ($1/\tau_{APD} \sim$ 50-100 Hz). This leads to an artificial broadening of approximately 1/10 of the typical inhomogeneous broadening of the molecules in the nanocrystal. The hybrid device is actuated electrically using a low-noise DC voltage source.

**References:**

[1] I. Aharonovich, D. Englund, and M. Toth, "Solid-state single-photon emitters," Nature Photonics **10**(10), 631–641 (2016).
[2] G. Kucsko, P.C. Maurer, N.Y. Yao, M. Kubo, H.J. Noh, P.K. Lo, H. Park, and M.D. Lukin, "Nanometre-scale thermometry in a living cell," Nature **500**(7460), 54–58 (2013).
[3] K.G. Schädler, C. Ciancico, S. Pazzagli, P. Lombardi, A. Bachtold, C. Toninelli, A. Reserbat-Plantey, and F.H.L. Koppens, "Electrical Control of Lifetime-Limited Quantum Emitters Using 2D Materials," Nano Lett. **19**(6), 3789–3795 (2019).
[4] D. Le Sage, K. Arai, D.R. Glenn, S.J. DeVience, L.M. Pham, L. Rahn-Lee, M.D. Lukin, A. Yacoby, A. Komeili, and R.L. Walsworth, "Optical magnetic imaging of living cells," Nature **496**(7446), 486–489 (2013).
[5] L. Gaudreau, K.J. Tielrooij, G.E.D.K. Prawiroatmodjo, J. Osmond, F.J.G. De Abajo, and F.H.L. Koppens, "Universal Distance-Scaling of Nonradiative Energy Transfer to Graphene," Nano Lett. **13**(5), 2030–2035 (2013).
[6] S. Kumar, A. Kaczmarczyk, and B.D. Gerardot, "Strain-Induced Spatial and Spectral Isolation of Quantum Emitters in Mono- and Bilayer WSe$_2$," Nano Lett. **15**(11), 7567–7573 (2015).
[7] C. Palacios-Berraquero, M. Barbone, D.M. Kara, X. Chen, I. Goykhman, D. Yoon, A.K. Ott, J. Beitner, K. Watanabe, T. Taniguchi, A.C. Ferrari, and M. Atatüre, "Atomically thin quantum light-emitting diodes," Nat Commun **7**(1), 12978 (2016).
[8] A. Branny, S. Kumar, R. Proux, and B.D. Gerardot, "Deterministic strain-induced arrays of quantum emitters in a two-dimensional semiconductor," Nat Commun **8**(1), 15053 (2017).
[9] J. Klein, A. Kuc, A. Nolinder, M. Altzschner, J. Wierzbowski, F. Sigger, F. Kreupl, J.J. Finley, U. Wurstbauer, A.W. Holleitner, and M. Kaniber, "Robust valley polarization of helium ion modified atomically thin MoS$_2$," 2D Mater. **5**(1), 011007 (2017).
[10] E. Mitterreiter, B. Schuler, K.A. Cochrane, U. Wurstbauer, A. Weber-Bargioni, C. Kastl, and A.W. Holleitner, "Atomistic Positioning of Defects in Helium Ion Treated Single-Layer MoS$_2$," Nano Lett. **20**(6), 4437–4444 (2020).
[11] K. Barthelmi, J. Klein, A. Hötger, L. Sigl, F. Sigger, E. Mitterreiter, S. Rey, S. Gyger, M. Lorke, M. Florian, F. Jahnke, T. Taniguchi, K. Watanabe, V. Zwiller, K.D. Jöns, U. Wurstbauer, C. Kastl, A. Weber-Bargioni, J.J. Finley, K. Müller, and A.W. Holleitner, "Atomistic defects as single-photon emitters in atomically thin MoS$_2$," Applied Physics Letters **117**(7), 070501 (2020).
[12] R. Smit, A. Tebyani, J. Hameury, S.J. van der Molen, and M. Orrit, "Sharp zero-phonon lines of single organic molecules on a hexagonal boron-nitride surface," Nat Commun **14**(1), 7960 (2023).
[13] S. Pazzagli, P. Lombardi, D. Martella, M. Colautti, B. Tiribilli, F.S. Cataliotti, and C. Toninelli, "Self-Assembled Nanocrystals of Polycyclic Aromatic Hydrocarbons Show Photostable Single-Photon Emission," ACS Nano **12**(5), 4295–4303 (2018).

[14] A.A.L. Nicolet, P. Bordat, C. Hofmann, M.A. Kol'chenko, B. Kozankiewicz, R. Brown, and M. Orrit, "Single Dibenzoterrylene Molecules in an Anthracene Crystal: Main Insertion Sites," ChemPhysChem **8**(13), 1929–1936 (2007).

[15] J. Tisler, T. Oeckinghaus, R.J. Stöhr, R. Kolesov, R. Reuter, F. Reinhard, and J. Wrachtrup, "Single Defect Center Scanning Near-Field Optical Microscopy on Graphene," Nano Lett. **13**(7), 3152–3156 (2013).

[16] S. Han, C. Qin, Y. Song, S. Dong, Y. Lei, S. Wang, X. Su, A. Wei, X. Li, G. Zhang, R. Chen, J. Hu, L. Xiao, and S. Jia, "Photostable fluorescent molecules on layered hexagonal boron nitride: Ideal single-photon sources at room temperature," The Journal of Chemical Physics **155**(24), 244301 (2021).

[17] J.-M. Caruge, and M. Orrit, "Probing local currents in semiconductors with single molecules," Phys. Rev. B **64**(20), 205202 (2001).

[18] A. Shkarin, D. Rattenbacher, J. Renger, S. Hönl, T. Utikal, P. Seidler, S. Götzinger, and V. Sandoghdar, "Nanoscopic Charge Fluctuations in a Gallium Phosphide Waveguide Measured by Single Molecules," Phys. Rev. Lett. **126**(13), 133602 (2021).

[19] D. Halbertal, M. Ben Shalom, A. Uri, K. Bagani, A.Y. Meltzer, I. Marcus, Y. Myasoedov, J. Birkbeck, L.S. Levitov, A.K. Geim, and E. Zeldov, "Imaging resonant dissipation from individual atomic defects in graphene," Science **358**(6368), 1303–1306 (2017).

[20] J. Martin, B.E. Feldman, R.T. Weitz, M.T. Allen, and A. Yacoby, "Local Compressibility Measurements of Correlated States in Suspended Bilayer Graphene," Phys. Rev. Lett. **105**(25), 256806 (2010).

[21] V. Puller, B. Lounis, and F. Pistolesi, "Single Molecule Detection of Nanomechanical Motion," Phys. Rev. Lett. **110**(12), 125501 (2013).

[22] A.A. Balandin, "Low-frequency 1/f noise in graphene devices," Nature Nanotech **8**(8), 549–555 (2013).

[23] M. Kamada, A. Laitinen, W. Zeng, M. Will, J. Sarkar, K. Tappura, H. Seppä, and P. Hakonen, "Electrical Low-Frequency $1/f^{\gamma}$ Noise Due to Surface Diffusion of Scatterers on an Ultra-low-Noise Graphene Platform," Nano Lett. **21**(18), 7637–7643 (2021).

[24] A. Moradi, Z. Ristanović, M. Orrit, I. Deperasińska, and B. Kozankiewicz, "Matrix-induced Linear Stark Effect of Single Dibenzoterrylene Molecules in 2,3-Dibromonaphthalene Crystal," Chemphyschem **20**(1), 55–61 (2019).

[25] Ch. Brunel, Ph. Tamarat, B. Lounis, J.C. Woehl, and M. Orrit, "Stark Effect on Single Molecules of Dibenzanthanthrene in a Naphthalene Crystal and in a *n*-Hexadecane Shpol'skii Matrix," J. Phys. Chem. A **103**(14), 2429–2434 (1999).

[26] M. Colautti, F.S. Piccioli, Z. Ristanović, P. Lombardi, A. Moradi, S. Adhikari, I. Deperasinska, B. Kozankiewicz, M. Orrit, and C. Toninelli, "Laser-Induced Frequency Tuning of Fourier-Limited Single-Molecule Emitters," ACS Nano **14**(10), 13584–13592 (2020).

[27] A.A.L. Nicolet, M.A. Kol'chenko, C. Hofmann, B. Kozankiewicz, and M. Orrit, "Nanoscale probing of charge transport in an organic field-effect transistor at cryogenic temperatures," Phys. Chem. Chem. Phys. **15**(12), 4415–4421 (2013).

[28] B. Terrés, L.A. Chizhova, F. Libisch, J. Peiro, D. Jörger, S. Engels, A. Girschik, K. Watanabe, T. Taniguchi, S.V. Rotkin, J. Burgdörfer, and C. Stampfer, "Size quantization of Dirac fermions in graphene constrictions," Nat Commun **7**(1), 11528 (2016).

[29] S. Grandi, K.D. Major, C. Polisseni, S. Boissier, A.S. Clark, and E.A. Hinds, "Quantum dynamics of a driven two-level molecule with variable dephasing," Phys. Rev. A **94**(6), 063839 (2016).

[30] W. Häfner, and W. Kiefer, "Raman spectroscopic investigations on molecular crystals: Pressure and temperature dependence of external phonons in naphthalene-d8 and anthracene-d1," The Journal of Chemical Physics **86**(8), 4582–4596 (1987).

[31] S. Somanchi, B. Terrés, J. Peiro, M. Staggenborg, K. Watanabe, T. Taniguchi, B. Beschoten, and C. Stampfer, "From Diffusive to Ballistic Transport in Etched Graphene Constrictions and Nanoribbons," Annalen Der Physik **529**(11), 1700082 (2017).

[32] M.J.H. Ku, T.X. Zhou, Q. Li, Y.J. Shin, J.K. Shi, C. Burch, L.E. Anderson, A.T. Pierce, Y. Xie, A. Hamo, U. Vool, H. Zhang, F. Casola, T. Taniguchi, K. Watanabe, M.M. Fogler, P. Kim, A. Yacoby, and R.L. Walsworth, "Imaging viscous flow of the Dirac fluid in graphene," Nature **583**(7817), 537–541 (2020).

[33] A.J. Healey, S.C. Scholten, T. Yang, J.A. Scott, G.J. Abrahams, I.O. Robertson, X.F. Hou, Y.F. Guo, S. Rahman, Y. Lu, M. Kianinia, I. Aharonovich, and J.-P. Tetienne, "Quantum microscopy with van der Waals heterostructures," Nat. Phys. **19**(1), 87–91 (2023).

[34] D.A. Bandurin, I. Torre, R.K. Kumar, M. Ben Shalom, A. Tomadin, A. Principi, G.H. Auton, E. Khestanova, K.S. Novoselov, I.V. Grigorieva, L.A. Ponomarenko, A.K. Geim, and M. Polini, "Negative local resistance caused by viscous electron backflow in graphene," Science **351**(6277), 1055–1058 (2016).

[35] Y. Xu, S. Liu, D.A. Rhodes, K. Watanabe, T. Taniguchi, J. Hone, V. Elser, K.F. Mak, and J. Shan, "Correlated insulating states at fractional fillings of moiré superlattices," Nature **587**(7833), 214–218 (2020).

[36] F. Pizzocchero, L. Gammelgaard, B.S. Jessen, J.M. Caridad, L. Wang, J. Hone, P. Bøggild, and T.J. Booth, "The hot pick-up technique for batch assembly of van der Waals heterostructures," Nat Comm. **7**(1), 11894 (2016).